\documentclass[%
 reprint,
 amsmath,amssymb,
 aps,
]{revtex4-2}

\usepackage{amsmath}
\usepackage{graphicx}
\usepackage{color}
\usepackage{dcolumn}
\usepackage{bm}
\usepackage[mathlines]{lineno}
\usepackage{array}
\usepackage{subfigure}

\begin{document}

\title{Search for the charged lepton flavor violating decay $J/\psi\to
  e\tau$}


\author{
\begin{small}
\begin{center}
M.~Ablikim$^{1}$, M.~N.~Achasov$^{10,c}$, P.~Adlarson$^{67}$, S. ~Ahmed$^{15}$, M.~Albrecht$^{4}$, R.~Aliberti$^{28}$, A.~Amoroso$^{66A,66C}$, M.~R.~An$^{32}$, Q.~An$^{63,49}$, X.~H.~Bai$^{57}$
% ...
\\
\vspace{0.2cm}
(BESIII Collaboration)\\
\vspace{0.2cm} {\it
$^{1}$ Institute of High Energy Physics, Beijing 100049, People's Republic of China\\
% ...
$^{4}$ Bochum Ruhr-University, D-44780 Bochum, Germany\\
% ...
$^{10}$ G.I. Budker Institute of Nuclear Physics SB RAS (BINP), Novosibirsk 630090, Russia\\
% ...
$^{15}$ Helmholtz Institute Mainz, Johann-Joachim-Becher-Weg 45, D-55099 Mainz, Germany\\
% ...
$^{28}$ Johannes Gutenberg University of Mainz, Johann-Joachim-Becher-Weg 45, D-55099 Mainz, Germany\\
% ...
$^{32}$ Liaoning Normal University, Dalian 116029, People's Republic of China\\
% ...
$^{49}$ State Key Laboratory of Particle Detection and Electronics, Beijing 100049, Hefei 230026, People's Republic of China\\
% ...
$^{57}$ University of Jinan, Jinan 250022, People's Republic of China\\
% ...
$^{63}$ University of Science and Technology of China, Hefei 230026, People's Republic of China\\
% ...
$^{66}$ University of Turin and INFN, (A)University of Turin, I-10125, Turin, Italy; (B)University of Eastern Piedmont, I-15121, Alessandria, Italy; (C)INFN, I-10125, Turin, Italy\\
$^{67}$ Uppsala University, Box 516, SE-75120 Uppsala, Sweden\\
% ...
$^{c}$ Also at the Novosibirsk State University, Novosibirsk, 630090, Russia\\
% ...
}
\vspace{0.2cm}
(Dated: Oct. 22, 2020)\\
\vspace{0.4cm} 
\end{center}
\end{small}
}


\begin{abstract}
  A search for the charged lepton flavor violating decay
  $J/\psi\to e^{\pm}\tau^{\mp}$ with $\tau^{\mp} \to \pi^{\mp}\pi^0\nu_{\tau}$ is performed with
  about $10\times10^9$  $J/ \psi$ events collected with the
  BESIII detector at the BEPCII. No significant signal is observed,
and an upper limit is set on the branching fraction
  $\mathcal{B}(J/\psi\to e^{\pm}\tau^{\mp})<7.5\times10^{-8}$ at the 90$\%$ confidence
  level. This improves the previously published limit by two orders of magnitude.
\end{abstract}

\maketitle


\section{\label{sec:level1}Introduction}
In the Standard Model (SM) of particle physics, the charged lepton
flavor violating (CLFV) process is forbidden~\cite{Marciano1, Marciano2}, therefore any significant sign of a signal could indicate physics beyond the SM. 
In recent years, there have been active phenomenological exploration of the sources of lepton flavor symmetry breaking.
Many physics models beyond the SM could allow CLFV processes to take place, such as
supersymmetry~\cite{Borzumati,Arbey,Paradisi,Calibbi}, the two Higgs doublet model~\cite{Branco,Crivellin},
and models including a fourth generation of quarks and leptons~\cite{Buras}.
The searches have been carried out in a variety of experimental endeavors. For example, the MEG collaboration searched for the
decay $\mu^+\to\gamma e^+$ and set the best upper limit (UL) of $\mathcal{B}(\mu^+\to\gamma e^+) < 4.2 \times 10^{-13}$~\cite{MEG} up to now,
while the $BABAR$ collaboration found a limit of $\mathcal{B}(\tau^+\to\gamma e^+) < 3.3 \times 10^{-8}$~\cite{BaBar}.
Meanwhile, many experiments searched for CLFV processes in the decays of pseudoscalar mesons, vector mesons, gauge bosons, and the Higgs boson,
e.g., pions~\cite{exp_pion}, kaons~\cite{exp_kaon}, $B$ mesons~\cite{exp_B_LHCb,exp_B_BaBar}, bottomonium states~\cite{exp_U_CLEO,exp_U_BaBar},
$Z^0$~\cite{exp_Z_ATLAS1,exp_Z_ATLAS2}, and Higgs~\cite{exp_H_CMS1,exp_H_CMS2}.

There are various theoretical predictions on CLFV in the charmonium states using model-independent methods~\cite{Zhangxm,TGutche2011},
unparticle physics~\cite{Sun}, and the minimal supersymmetric model with gauged baryon number and lepton number~\cite{Dong}, etc.
Some of these predictions constrain $\mathcal{B}(J/\psi\to e^{\pm}\mu^{\mp})$ to the order of $10^{-13}$,
while $\mathcal{B}(J/\psi\to e^{\pm}\tau^{\mp} )$ and $\mathcal{B}(J/\psi\to \mu^{\pm}\tau^{\mp} )$ to $10^{-9}$.
With $58\times10^6$ $J/\psi$ events, the BES collaboration obtained experimental ULs of various decays of charmonium states, namely
$\mathcal{B}(J/\psi\to e^{\pm}\mu^{\mp}$) $<1.1\times10^{-6}$~\cite{BESII_emu}, $\mathcal{B}(J/\psi\to e^{\pm}\tau^{\mp})$ $< 8.3 \times 10^{-6}$, and $\mathcal{B}(J/\psi\to \mu^{\pm}\tau^{\mp})$ $< 2.0 \times 10^{-6}$~\cite{etau}.
More recently and based on $225\times10^6$ $J/\psi$ events collected with BESIII, an UL of $\mathcal{B}(J/\psi\to e^{\pm}\mu^{\mp})$ $< 1.6 \times 10^{-7}$ was obtained~\cite{BESIII_emu}.

In this paper, the CLFV process of $J/\psi\to e^{\pm}\tau^{\mp}$ with $\tau^{\mp} \to \pi^{\mp}\pi^0\nu_{\tau}$ is probed based on $10\times10^9$ $J/ \psi$ events collected with the BESIII detector. A semiblind analysis is performed to avoid a possible bias. About $10\%$ of the full data sample are randomly selected.
Besides the selected data, several simulation samples, and independent continuum data samples are used to optimize the event selection criteria, study the background,
and estimate the systematic uncertainties. The final results are obtained with the full data sample by repeating the validated analysis strategy. In the rest of this paper, the charge conjugated channel is implied unless otherwise specified.

\section{\label{sec:level2}BESIII Detector}
The BESIII detector is a magnetic
spectrometer~\cite{Ablikim:2009aa} located at the Beijing Electron
Positron Collider (BEPCII)~\cite{Yu:IPAC2016-TUYA01,Ablikim:2019hff}. The
cylindrical core of the BESIII detector consists of a helium-based
 multilayer drift chamber (MDC), a plastic scintillator time-of-flight
system (TOF), and a CsI(Tl) electromagnetic calorimeter (EMC),
which are all enclosed in a superconducting solenoidal magnet
providing a $1.0$~T($0.9$~T in 2012) magnetic field. The solenoid is supported by an
octagonal flux-return yoke with resistive plate counter muon
identifier modules interleaved with steel. The acceptance of
charged particles and photons is $93\%$ over $4\pi$ solid angle. The
charged-particle momentum resolution at $1~{\rm GeV}/c$ is
$0.5\%$, and the ${\rm d}E/{\rm d}x$ resolution is $6\%$ for the electrons
from Bhabha scattering. The EMC measures photon energies with a
resolution of $2.5\%$($5\%$) at $1$~GeV in the barrel (end cap)
region. The time resolution of the TOF barrel part is $68$~ps, while
that of the end cap part is $110$~ps. The end cap TOF system is upgraded in 2015 with a multigap resistive plate chamber technology,
providing a time resolution of $60$~ps~\cite{etof}.

\section{\label{sec:level3}Data samples and Monte Carlo Simulation}
The analysis is based on $J/\psi$ events collected in the years of 2009, 2012, 2018 and 2019 at BESIII. 
The total number of $J/\psi$ events collected in these years is determined using inclusive $J/\psi$ decays with the method described in Ref.~\cite{Njpsi}. For the selected inclusive $J/\psi$ events, the background due to QED processes, beam-gas interactions and cosmic rays is estimated using the continuum data samples at $\sqrt{s} = 3.08$ GeV. The detection efficiency for the inclusive $J/\psi$ decays is obtained using the experimental data sample of $\psi(3686) \rightarrow \pi^+\pi^-J/\psi$. The efficiency difference between the $J/\psi$ produced at rest and the $J/\psi$ from the decay $\psi(3686)\rightarrow\pi^+\pi^-J/\psi$ is estimated by comparing the corresponding efficiencies in a Monte Carlo (MC) simulation. The uncertainties related to the signal MC model, track reconstruction efficiency, fit to the $J/\psi$ mass peak, background estimation, noise mixing and reconstruction efficiency for the pions recoiling against the $J/\psi$ are studied. Finally, the number of $J/\psi$ events collected at BESIII is determined to be $N_{J/\psi}= (10087\pm44)\times10^{6}$. Among them, in 2009 and 2012, the total $J/\psi$ number is $(1310.6\pm7.0)\times10^{6}$~\cite{Njpsi} and this data sample is denoted as ``data sample I.'' Likewise, the data sample collected in 2018 and 2019 is denoted as ``data sample II.''

MC simulated samples produced with the {\sc
geant4}-based~\cite{geant4} package, which
includes the geometric description of the BESIII detector and the
detector response, are used to determine the detection efficiency
and to estimate the backgrounds. The simulation takes the beam-energy
spread and initial-state radiation in the $e^+e^-$
annihilations into account, modeled with the generator
{\sc kkmc}~\cite{ref:kkmc}. The inclusive MC sample consists of the production
of the $J/\psi$ resonance and the continuum processes incorporated in {\sc
kkmc}~\cite{ref:kkmc}.  The inclusive MC sample contains $1.225 \times 10^9$ $J/\psi$ events for data sample I and $8.700 \times 10^9$ $J/\psi$ events for data sample II. The known decay modes are modeled with {\sc
evtgen}~\cite{ref:evtgen} using branching fractions (BFs) taken from the
Particle Data Group (PDG)~\cite{PDG2020}, and the remaining unknown decays
from the $J/\psi$ with {\sc
lundcharm}~\cite{ref:lundcharm}. The final-state radiations (FSR)
from charged final-state particles are incorporated with the {\sc
  photos} package~\cite{photos}. The VLL model, which describes the decay of a vector meson to two charged leptons~\cite{ref:evtgen},
is used to generate the process $J/ \psi \rightarrow e^{-}\tau^{+}$. The TAUHADNU generator, which describes the $\tau$ semileptonic decay with several pions~\cite{ref:evtgen},
is used to generate the process $\tau^{+}\to \pi^{+}\pi^0\nu$. This generator is based on conserved vector currents and a chiral Lagrangian model~\cite{TAUHADNU1}
with parameters taken from the paper of the CLEO collaboration~\cite{TAUHADNU2}. The MC-generated samples simulating signal events,
in the following abbreviated as signal MC samples, are generated individually for each data sample and denoted as ``signal sample I'' and ``signal sample II''. To study background contributions, many potential backgrounds of $J/\psi$ decays are generated exclusively with a much larger statistics than each data sample, such as $J/\psi \to  \pi^+\pi^-\pi^0$ with Dalitz amplitudes,
$J/\psi \to  \rho\pi$ with helicity amplitudes, as well as $J/\psi \to  \omega f_2(1270)$ and $ \bar{p}n\pi^+ $ with phase space distributions.

\section{\label{sec:level4}Event Selection}
Two charged particles with zero net charge are required to satisfy the polar-angle condition $|\cos \theta| < 0.8$ with respect to the beam axis.
Their closest approaches to the interaction point are required to be within $10$~cm in the beam direction and within $1$~cm in the plane perpendicular to the beam.
The particle identification (PID) is performed by combining the energy-loss measurement, ${\rm d}E/{\rm d}x$, obtained from the MDC and the time-of-flight information
from the TOF. PID confidence levels (C.L.) are calculated for the electron ($CL_{e}$), pion ($CL_\mathrm{\pi}$), proton ($CL_\mathrm{p}$), and kaon ($CL_\mathrm{K}$) hypotheses.
The electron (pion) candidate requires the electron (pion) hypothesis to have the highest PID confidence levels among the four hypotheses. For electron candidates,
the $CL_{e}$/($CL_{e}$ + $CL_\mathrm{\pi}$) ratio is required to be larger than $0.95$ and the $E/p$ is larger than 
$0.8$ to further improve the electron identification. Here, the variables $E$ and $p$ refer to the energy deposition of the charged track in the EMC and
its momentum measured with the MDC, respectively.

Electromagnetic showers in the EMC are identified as photon candidates only if the following criteria are satisfied. The energy deposition is required to be larger than 25~MeV
in the barrel ($|\cos \theta| < 0.8$) region and 50~MeV in the end cap ($0.86 < |\cos \theta| < 0.92$) regions. To eliminate showers produced by charged particles,
the photon candidates are required to be separated from the extrapolated positions of any charged track by more than $10^\circ$. To suppress electronic noise and unrelated energy depositions,
the EMC time deviation from the event start time is required to be within $700$~ns. At least two photons satisfying these selection criteria are required in the final state.
The $\pi^0$ candidate is reconstructed from photon pairs whose invariant mass $M(\gamma\gamma)$ is required to satisfy $115$ MeV/$c^2$ $< M(\gamma\gamma) <$ $150$ MeV/$c^2$. To improve the momentum resolution, a kinematic fit is applied by constraining the two-photon invariant mass to the nominal $\pi^0$ mass, and the refined four momenta of the photons are used for further analysis.

\begin{figure}[htb]
\centering
\subfigure{
\begin{minipage}[b]{0.45\textwidth}
\includegraphics[width=1\textwidth]{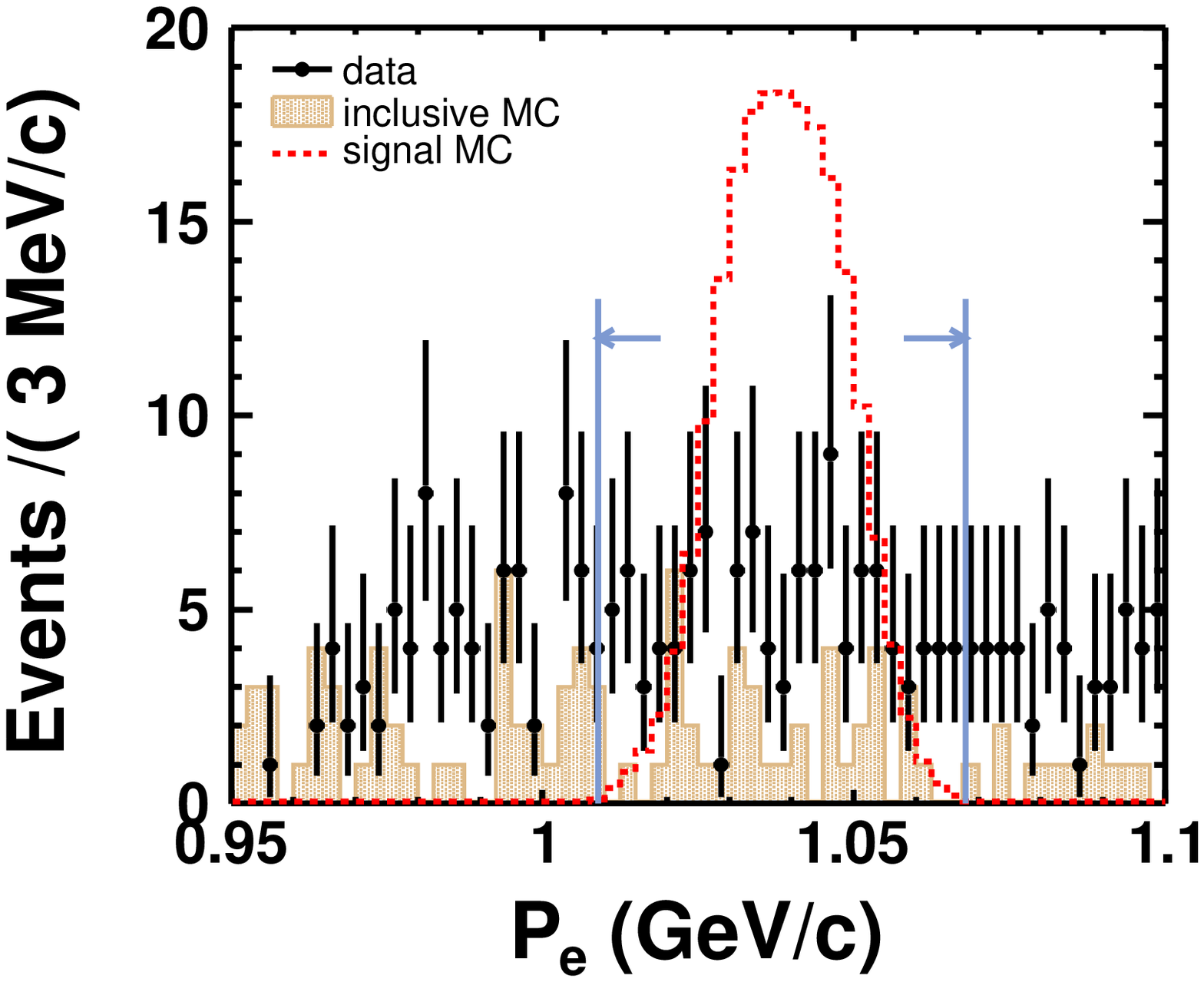}\put(-65,135){\bf  ~(a)}
\end{minipage}
}
\subfigure{
\begin{minipage}[b]{0.45\textwidth}
\includegraphics[width=1\textwidth]{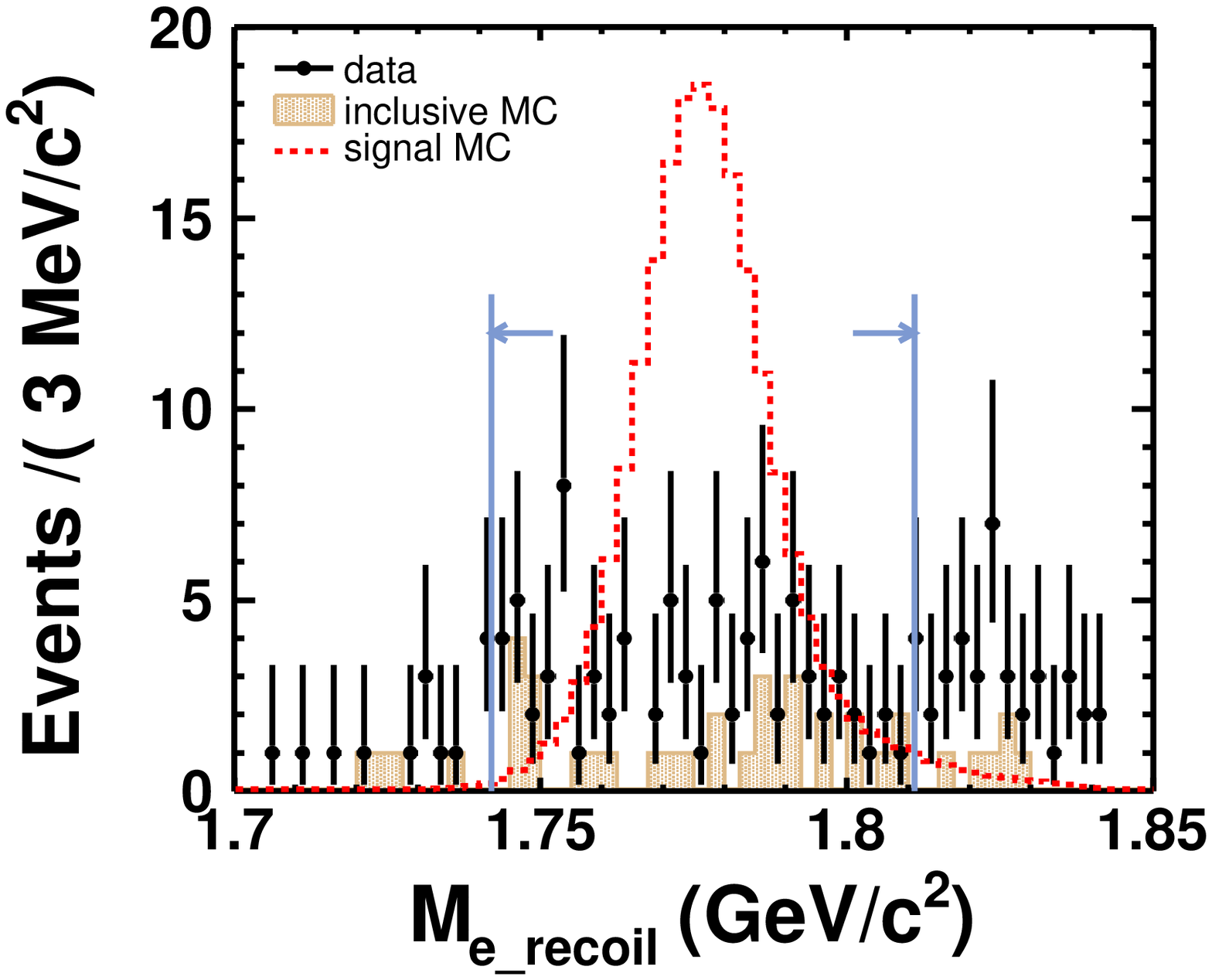}\put(-65,135){\bf  ~(b)}
\end{minipage}
}
\caption{(a) The electron-momentum $P_{e}$ and (b) recoil-mass $M_{e\_\mathrm{recoil}}$ distributions. The dots with error bars are data, while the shaded histograms are obtained from a $J/\psi$ inclusive MC sample normalized to data. The dashed lines show the arbitrarily scaled signal shape extracted from the MC signal samples. The areas between the arrows indicate the selection region.}
\label{Pe_mtau}
\end{figure}

The final-state electron from the process $J/\psi\to e^{-}\tau^{+}$ is monochromatic, therefore the momentum of the electron $P_{e}$ and the recoiling mass against the
electron $M_{e\_\mathrm{recoil}}$ are required to be within $1.009$ GeV/$c$ $< P_{e} <$ $1.068$ GeV/$c$ and $1.742$ GeV/$c^2$ $< M_{e\_\mathrm{recoil}} <$ $1.811$ GeV/$c^2$, respectively.
The momentum and recoil-mass resolutions are found, using MC simulations, to be $0.010$ and $0.011$ GeV/$c^2$, respectively.
Figure~\ref{Pe_mtau} compares the momentum and recoil-mass distributions of the complete data sample with the corresponding signal MC simulation and $J/\psi$ inclusive MC samples. The possible background from continuum process would be discussed in the next section.

The missing energy $E_{\mathrm{miss}}$ is calculated by $E_\mathrm{miss}=E_\mathrm{CMS}-E_{e}-E_{\pi}-E_{\pi^0}$, where $E_\mathrm{CMS}$ is the center-of-mass energy of the
initial $e^{+}e^{-}$ system, while $E_{e}$, $E_{\pi}$, and $E_{\pi^0}$ are the energies of the electron, charged pion, and neutral pion in the rest frame of the $e^{+}e^{-}$ system.
The $E_\mathrm{miss}$ is required to be larger than $0.43$~GeV to suppress the background events whose final states are all detected.
The variable $U_{\mathrm{miss}}$, calculated by $U_{\mathrm{miss}} = E_{\mathrm{miss}} - c|\vec{P}_{\mathrm{miss}}|$, is used to define the signal region to have a better resolution than the missing mass.
The variable $\vec{P}_\mathrm{miss}=\vec{P}_{J/\psi}-\vec{P}_{e}-\vec{P}_{\pi}-\vec{P}_{\pi^0}$ is the missing momentum, where $\vec{P}$ are the corresponding momenta in the rest frame of the $e^{+}e^{-}$ system of the particles indicated by the subscript. As the signal events peak near zero with one undetected neutrino, the signal region is defined to be $-0.081$ GeV $< U_\mathrm{miss} < $ $0.112$ GeV, which corresponds
to three standard deviations of the expected width determined from the signal MC sample.
The number of signal candidates for each data sample, $N_\mathrm{obs}$, is obtained by counting the number of entries that fall within the signal region.
After applying the above selection criteria, the detection efficiency of the signal sample I (II) is determined to be ($20.24 \pm 0.05$)$\%$ (($19.37 \pm 0.02$)$\%$).

\section{\label{sec:level5}Background study}
The dominant background contaminations stem from the continuum process (e.g. radiative Bhabha) and from hadronic $J/\psi$
decays such as $J/\psi \to  \pi^+\pi^-\pi^0$. 

The continuum background is studied with a $150$~pb$^{-1}$ data sample collected at $\sqrt s$ = $3.08$~GeV and a $2.93$~fb$^{-1}$ data sample taken at
$\sqrt s$ = $3.773$ GeV. The survived events are dominated by radiative Bhabha process, therefore the normalized background events from the continuum
processes are estimated with the assumption of a $1/s$ dependence of the cross section. Radiative Bhabha MC samples at different energy points are used to evaluate the uncertainty of this assumption to be about $24\%$. Single electron MC samples are used to study the electron momentum resolution differences at different energy points. The resolution differences are applied to radiative Bhabha MC samples, and the result shows this influence could be negligible in this study. The continuum background events are estimated to be $5.8 \pm1.8$ ($37.9 \pm 11.5$) for data sample I (II) with the uncertainties of statistics and the assumption of $1/s$ dependence taken into consideration.

The $J/\psi$ decay background is studied with the inclusive MC samples, and only a few events survive. Main background processes from $J/\psi \to  \pi^+\pi^-\pi^0$, $J/\psi \to  \rho\pi$,
$J/\psi \to  \omega f_2(1270)$, and $J/\psi \to \bar{p}n\pi^+ $ are studied with exclusive MC samples. The uncertainty in the $J/\psi$ decay modeling is determined to be about $16\%$ from the inclusive MC samples with and without {\sc lundcharm} model. The normalized background events from the $J/\psi$ decays are estimated to be $1.1 \pm 0.8$ ($25.7 \pm 6.4$) for data sample I (II) with statistical and $J/\psi$ decay modeling uncertainties taken into consideration.
The possible cross feed from the CLFV process $J/\psi \rightarrow e^{\pm}\tau^{\mp}$ whereby
the $\tau$ decays to other modes has been studied using a $\tau$ inclusive MC sample modeled by {\sc evtgen}~\cite{ref:evtgen} and found to be negligible ($0.3\%$). 
The background events from $J/\psi$ decay processes are normalized according to the BFs from the PDG~\cite{PDG2020},
the number of $J/\psi$ events, and the detection efficiencies determined from the exclusive MC samples. 

\begin{figure}[!http]
\centering
\includegraphics[width=0.5\textwidth]{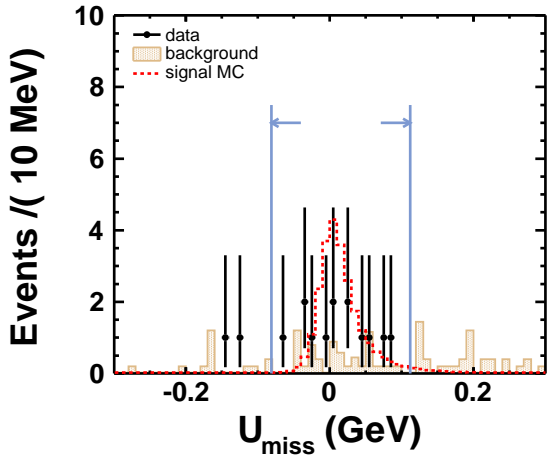}
\caption{The $U_\mathrm{miss}$ distribution of data sample I and corresponding background. The dots with error bars are data, while the
  shaded histogram is from the normalized continuum sample as well as the $J/\psi$ inclusive MC samples. The dashed line shows the arbitrarily scaled signal MC shape extracted
  from the signal sample I. The areas between the arrows represent the signal region.}
\label{unblind1}
\end{figure}

\begin{figure}[!http]
\centering
\includegraphics[width=0.5\textwidth]{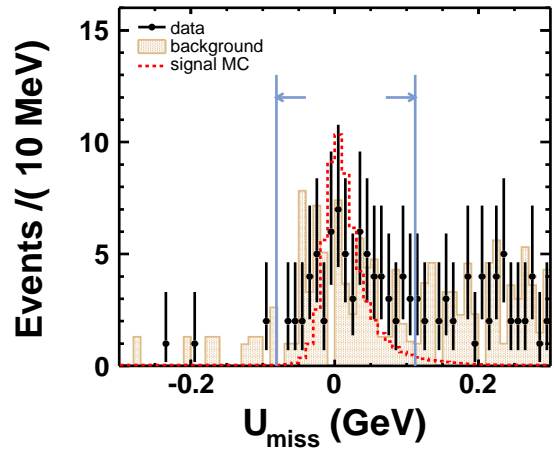}
\caption{Same as Fig.~\ref{unblind1} except for data sample II and signal sample II.}
\label{unblind2}
\end{figure}

The normalized background events from continuum processes and $J/\psi$ decay processes discussed above are utilized to estimate the number of
background events left in the signal region. In total, $6.9 \pm 1.9$ ($63.6 \pm 13.2$) background events are expected for the data sample I (II).
Some background events with additional soft tracks contribute near the $U_\mathrm{miss} = 0$ region. The signal region is opened after completing
the optimization of the analysis algorithms and the background study. Figures~\ref{unblind1} and \ref{unblind2} depict
$U_\mathrm{miss}$ for the data samples I and II, respectively.

\section{\label{sec:level6}Systematic Uncertainties}
Systematic uncertainties mainly come from uncertainties in the total number of $J/\psi$ decays, the quoted intermediate BF, the background estimation, and in efficiencies associated with signal modeling, PID and tracking of charged particles, the photon detection, the $\pi^0$ reconstruction, and kinematic variable requirements. The details of most of sources are described below, while the uncertainties from background estimation have been considered in the Sec.~\ref{sec:level5}.

The uncertainty in the number of $J/\psi$ events is determined to be
0.5$\%$ for the data sample I~\cite{Njpsi}, and $0.4\%$ for the data sample II. 
The uncertainty in the quoted BF of $\tau^{+} \to \pi^{-}\pi^0\nu_{\tau}$ is $0.4\%$~\cite{PDG2020}. To estimate the uncertainty in the signal MC model,the generator producing $\tau^{+} \to \pi^{-}\pi^0\nu_{\tau}$ decays is changed to the TAUVECTORNU generator (for $\tau^+ \to \rho^+\nu_{\tau}$) and the
VSS generator (for $\rho^+ \to \pi^+\pi^0$)~\cite{ref:evtgen}. The TAUVECTORNU generator simulates the decay of a $\tau$ lepton into a vector
particle and a neutrino, while the VSS generator simulates the decay a vector meson into a pair of scalar particles.
The relative change in the detection efficiency of signal sample I, $0.6\%$, is assigned as the uncertainty. The relative change
in the efficiency for sample II is found to be negligible.

The uncertainty in the PID of pions is $1.0\%$ per charged pion, as determined from a study of the control sample of the process
$J/\psi\to\rho\pi$~\cite{sys_pid_pi}. The MDC tracking efficiency of charged pions is studied using the control sample of $J/\psi\to\pi^+\pi^-p\bar{p}$ decays,
and the difference between the data and MC simulation is $1.0\%$ for each charged pion~\cite{sys_trk_pi}.
The PID and tracking efficiencies of electrons are obtained from a control sample of radiative Bhabha scattering
$e^+e^- \to \gamma e^+e^-$ (including $J/\psi \to \gamma e^+e^-$) corresponding to
the center-of-mass energy of the $J/\psi$ resonance.
For the electron-PID study, the same PID requirements as applied to the dataset of interest are exposed to the control samples.
Similarly, for the electron-tracking study, we applied the same conditions for the polar angle and for the closest distance to the interaction point
as was used for the data of interest.
Differences in PID (tracking) efficiencies between the data and MC simulations are obtained for each bin of a two-dimensional distribution representing
the momentum (transverse momentum) versus the polar angle of the electron tracks. These results are subsequently used to determine
the overall weighted differences per track for PID (tracking). We obtained PID and tracking uncertainties for electrons of the
signal sample I (II) of $0.4\%$ ($0.9\%$) and $0.1\%$ ($0.1\%$) per track, respectively. 

The photon detection efficiency is studied with the control sample based on
$J/\psi \to \pi^+\pi^-\pi^0$, $\pi^0\to\gamma\gamma$
events~\cite{sys_pid_pi}. The difference between data and MC simulation is $0.5\%$ ($1.5\%$) for a photon in the EMC barrel (end cap) region.
The average difference, $0.5\%$ per photon, is taken as systematic uncertainty. The
total systematic uncertainty due to uncertainties in the photon-detection efficiency is estimated to be $1.0\%$.
The uncertainty related to the $\pi^0$ reconstruction is determined to be
$1.0\%$ for the two samples using a $J/\psi\to\pi^+\pi^-\pi^0$ control sample as described in Ref.~\cite{sys_rec_pi0}.

The systematic uncertainties related to $P_{e}$ and $M_{e\_\mathrm{recoil}}$ requirements are studied with the control sample of
the process $e^+e^- \to \gamma e^+e^-$ (including $J/\psi \to \gamma e^+e^-$) at the center-of-mass energy of the $J/\psi$ resonance.
The differences in efficiency between the data and MC simulation for these two kinematic variables are studied by varying the event-selection requirement ranges
while taking into account the correlation between them. This uncertainty is determined to be $3.0\%$ ($3.3\%$) for sample I (II). The same control sample is used to study the uncertainty
associated with $E_\mathrm{miss}$ requirement. The electron with the lowest momentum is assumed to be a missing track, and the data-MC differences of the resulting missing energy are derived as correction factors to be applied to the $E_\mathrm{miss}$ distribution
of the signal MC sample. Then the difference in efficiency, $1.0\%$ ($0.8\%$), between the signal MC sample with and without the correction is taken as the systematic
uncertainty for data sample I (II).

Table~\ref{systematic} summarizes all sources of systematic
uncertainties discussed above. The total systematic uncertainties of each data sample are obtained by adding these uncertainties in quadrature.

\begin{table}[h]
\caption{Summary of the sources of systematic uncertainties and their estimated magnitudes. The correlated sources are marked with an asterisk,  which are added linearly when combining the two data samples. The negligible uncertainty is marked with a dash line.}
\begin{center}
\renewcommand\arraystretch{1.3}
\begin{tabular}{c|c|c }
\hline
\hline
Sources & sample I & sample II\\
\hline
Number of $J/\psi$ &0.5$\%$&0.4$\%$\\
Quoted BF* &0.4$\%$&0.4$\%$ \\
MC model &0.6$\%$ &-\\
Pion PID*&1.0$\%$&1.0$\%$\\
Pion tracking*&1.0$\%$&1.0$\%$ \\
Electron PID&0.4$\%$&0.9$\%$\\
Electron tracking*&0.1$\%$&0.1$\%$ \\
Photon detection* & 1.0$\%$& 1.0$\%$\\
$\pi^0$ reconstruction* &1.0$\%$&1.0$\%$  \\
$P_{e}$ and $M_{e\_\mathrm{recoil}}$ requirements  &3.0$\%$&3.3$\%$ \\
$E_\mathrm{miss}$ requirement  &1.0$\%$&0.8$\%$ \\

\hline
Total uncertainty &3.9$\%$&4.1$\%$ \\
\hline
\hline
\end{tabular}
\end{center}
\label{systematic}
\end{table}

\section{\label{sec:level7}Results}
Table~\ref{result} summarizes the extracted parameters of each sample. The parameters $N_\mathrm{bkg}^\mathrm{exp}$ and $\sigma_\mathrm{bkg}^\mathrm{exp}$
are the expected number of background events and its uncertainty in the signal region determined from the background study; $\epsilon_\mathrm{eff}^\mathrm{mc}$ and $\sigma_\mathrm{eff}^\mathrm{mc}$ denotes the efficiency and its uncertainty determined from signal MC samples and the study of systematic uncertainties.

\begin{table}[htbp]
\caption{A summary of the analysis results. See the text for details.}
\begin{center}
\renewcommand\arraystretch{1.3}
\begin{tabular}{c|c|c }
\hline
\hline
Results &sample I &sample II\\
\hline
$N_\mathrm{obs}$ &13 & 69\\
$N_\mathrm{bkg}^\mathrm{exp}$ &6.9&63.6\\
$\sigma_\mathrm{bkg}^\mathrm{exp}$ &1.9&13.2\\
$\epsilon_\mathrm{eff}^\mathrm{mc}$  &20.24$\%$  &19.37$\%$\\
$\sigma_\mathrm{eff}^\mathrm{mc}$ &0.79$\%$ &0.79$\%$\\
\hline
BF (90$\%$C.L.) &\multicolumn{2}{c}{$7.5\times10^{-8}$}\\
\hline
\hline
\end{tabular}
\end{center}
\label{result}
\end{table}

Since no significant signal is observed, a maximum likelihood estimator, extended from the profile-likelihood approach~\cite{TRolke}, is used to determine the UL on the BF of
$J/\psi\to e^{\pm}\tau^{\mp}$. The likelihood function of each sample which depends on the parameter of interest $\mathcal{B}(J/\psi\to e^{\pm}\tau^{\mp})$
and the nuisance parameters $\boldsymbol{\theta}$ = $(\epsilon_\mathrm{eff}, N_\mathrm{bkg})$ is defined as

\begin{small}
\begin{equation}\nonumber
\begin{aligned}
&\mathcal{L}(\mathcal{B}(J/\psi\to e^{\pm}\tau^{\mp}),\boldsymbol{\theta})=\\
&P(N_\mathrm{obs},\mathcal{B}(J/\psi\to e^{\pm}\tau^{\mp}) \cdot N_{J/\psi}\cdot\mathcal{B}_{\tau^{\mp}\to\pi^{\mp}\pi^0\nu_{\tau}} \cdot \epsilon_\mathrm{eff}+N_\mathrm{bkg})\\
&\cdot G(\epsilon_\mathrm{eff}^\mathrm{mc},\epsilon_\mathrm{eff},\sigma_\mathrm{eff}^\mathrm{mc})\cdot G(N_\mathrm{bkg}^\mathrm{exp},N_\mathrm{bkg},\sigma_\mathrm{bkg}^\mathrm{exp}),
\end{aligned}
\end{equation}
\end{small}
where the observed events are assumed to follow a Poisson distribution ($P$), while the detection efficiency $\epsilon_\mathrm{eff}$ and the background number $N_\mathrm{bkg}$ follow Gaussian distributions ($G$); $N_{J/\psi}$ is the number of $J/\psi$ events. 

\begin{figure}[!http]
\centering
\includegraphics[width=0.5\textwidth]{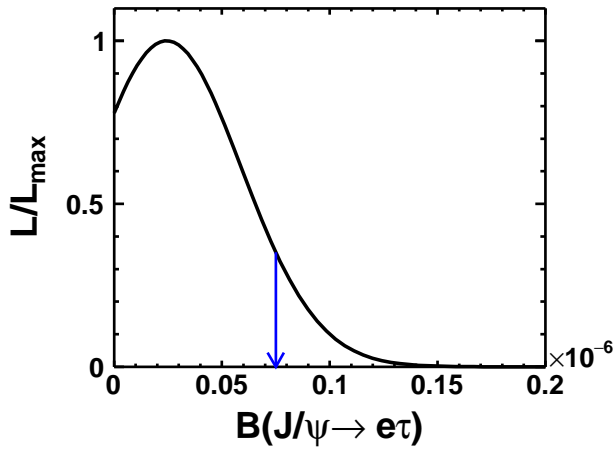}
\caption{The combined likelihood distribution as a function of the branching fraction of the data samples. The arrow points to the position of the UL at 90$\%$ C.L.}
\label{com_llh}
\end{figure}

The likelihood function is treated as the probability function, and the UL on the $J/\psi\to e^{\pm}\tau^{\mp}$ at 90$\%$ C.L. is determined by integrating the likelihood distribution in the physical region of $\mathcal{B}\ge 0$ based on the Bayesian method with the {\sc roostats} package~\cite{RooStats}. The combined likelihood distribution as a function of the BF from the data samples is shown in Fig.~\ref{com_llh}. The resultant UL is $\mathcal{B}(J/\psi\to e^{\pm}\tau^{\mp})<7.5\times10^{-8}$ (90$\%$ C.L.), where the detection efficiency, statistical and systematic uncertainties as well as the background estimation are all incorporated.

\section{\label{sec:level8}Summary}
This paper presents a search of the CLFV process  $J/\psi\to e^{\pm}\tau^{\mp}$ with $\tau^{\mp} \to \pi^{\mp}\pi^0\nu_{\tau}$ using a data sample based upon
$10\times10^9$ $J/ \psi$ events collected with the BESIII detector. A semiblind analysis found no significant excess in the datasets with
respect to the expected background. The UL is determined to be 
$\mathcal{B}(J/\psi\to e^{\pm}\tau^{\mp})<7.5\times10^{-8}$ (90$\%$ C.L.), where uncertainties are taken into account. This  improves the previous published limits~\cite{etau} by more than two orders of magnitude and can be used to constrain new physics parameter spaces. 

\begin{acknowledgments}

The BESIII collaboration thanks the staff of BEPCII and the IHEP computing center for their strong support. This
work is supported in part by National Key Research and Development Program of China under Contracts No.
2020YFA0406400, and No. 2020YFA0406300; National Natural Science Foundation of China (NSFC) under Contracts No. 11625523, No. 11635010, No. 11735014, No. 11822506, No. 11835012, No. 11935015, No. 11935016, No.
11935018, No. 11961141012, No. 12022510, No. 12025502, No. 12035009, No. 12035013,  and No. 12061131003;
the Chinese Academy of Sciences (CAS) Large-Scale Scientific Facility Program; Joint Large-Scale Scientific Facility
Funds of the NSFC and CAS under Contracts No. U1732263, and No. U1832207; CAS Key Research Program of Frontier Sciences under Contract No. QYZDJ-SSW-SLH040; 100 Talents Program of CAS; INPAC and Shanghai
Key Laboratory for Particle Physics and Cosmology; ERC under Contract No. 758462; European Union Horizon 2020
research and innovation programme under Contract No. Marie Sklodowska-Curie grant agreement Grant Agreement
No 894790; German Research Foundation DFG under Contracts No. 443159800, Collaborative Research Center No. CRC 1044, No. FOR 2359, No. FOR 2359, No. GRK 214; Istituto Nazionale di Fisica Nucleare, Italy;
Ministry of Development of Turkey under Contract No. DPT2006K-120470; National Science and Technology fund;
Olle Engkvist Foundation under Contract No. 200-0605; STFC (United Kingdom); The Knut and Alice Wallenberg
Foundation (Sweden) under Contract No. 2016.0157; The Royal Society, UK under Contracts No. DH140054, and No. DH160214; The Swedish Research Council; U. S. Department of Energy under Contracts No. DE-FG02-05ER41374, and No. DE-SC-0012069.

\end{acknowledgments}

%

\begin{thebibliography}{99}
\bibitem{Marciano1}
W.~J.~Marciano and A.~Sanda, 
Phys.\ Let. {\bf 67B}, 303 (1977) .

\bibitem{Marciano2}
W.~J.~Marciano,  T.~Mori, and J.~M.~Roney,
Annu.\ Rev.\  Nucl.\ Part. Sci. {\bf 58}, 315 (2008).

\bibitem{Borzumati}
F. Borzumati and A. Masiero, 
Phys.\ Rev.\ Lett. {\bf 57}, 961 (1986).

\bibitem{Arbey}
A.~Arbey, M.~Battaglia, A.~Djouadi, F.~Mahmoudi, and J.~Quevilon,
Phys.\ Lett.\ B {\bf 708}, 162 (2012).

\bibitem{Paradisi}
P.~Paradisi, 
J.\ High Energy Phys. 10 (2005) 006  .

\bibitem{Calibbi}
L.~Calibbi, P.~Paradisi, and R.~Ziegler, 
Eur. Phys. J. C {\bf 74}, 3211 (2014).

\bibitem{Branco}
G.~C.~Branco, P.~M.~Ferreira, L.~Lavoura, M,~N.~Rebelo, M.~Sher, and J.~P.~Silva,
Phys. Rep. {\bf 516}, 1 (2012). 

\bibitem{Crivellin}
A.~Crivellin, C.~Greub, and A.~Kokulu, 
Phys. Rev. D {\bf 87},  094031 (2013).

\bibitem{Buras}
A.~J.~Buras, B.~Duling, T.~Feldmann, T.~Heidsieck, and C.Promberger,
J.\ High Energy Phys. 09 (2010) 104.

\bibitem{MEG}
A.~ M.~Baldini {\it et al.} (MEG Collaboration),
Eur.\ Phys.\ J.\ C {\bf 76}, 434 (2016).

\bibitem{BaBar}
B.~Aubert {\it et al.} ($BABAR$ Collaboration),
Phys.\ Rev.\ Lett.\ {\bf 104}, 021802 (2010).

\bibitem{exp_pion}
D.~Ambrose {\it et al.} (BNL Collaboration),
Phys.\ Rev.\ Lett.\ {\bf 81}, 5734 (1998) .

\bibitem{exp_kaon}
E.~Abouzaid {\it et al.} (KTeV Collaboration), 
Phys.\ Rev.\ Lett.\ {\bf 100}, 131803 (2008).

\bibitem{exp_B_LHCb}
R.~Aaij {\it et al.} (LHCb Collaboration), 
J.\ High Energy Phys. 03 (2018) 078 .

\bibitem{exp_B_BaBar}
B.~Aubert {\it et al.} ($BABAR$ Collaboration), 
Phys.\ Rev.\ D {\bf 77} 091104 (2008).

\bibitem{exp_U_CLEO}
W.~Love {\it et al.} (CLEO Collaboration), 
Phys.\ Rev.\ Lett.\ {\bf 101}, 201601 (2008).

\bibitem{exp_U_BaBar}
J.~Lees \textit{et al.} ($BABAR$ Collaboration), 
Phys.\ Rev.\ Lett.\ \textbf{104}, 151802 (2010).

\bibitem{exp_Z_ATLAS1}
G.~Aad {\it et al.} (ATLAS Collaboration),
Phys.\ Rev.\ D {\bf 90}, 072010 (2014).

\bibitem{exp_Z_ATLAS2}
M.~Aaboud {\it et al.} (ATLAS Collaboration),
Phys.\ Rev.\ D {\bf 98}, 092010 (2018).

\bibitem{exp_H_CMS1}
V.~Khachatryan et al. (CMS Collaboration), 
Phys.\ Lett.\ B {\bf 763}, 472 (2016).

\bibitem{exp_H_CMS2}
A.~M.~Sirunyan \textit{et al.} (CMS Collaboration), 
J.\ High Energy Phys. 06 (2018) 001.

\bibitem{Zhangxm}
S.~Nussinov, R.~D.~Peccei, X.~M.~Zhang, 
Phys. Rev. D {\bf 63}, 016003 (2000). 

 
\bibitem{TGutche2011}
T.~Gutsche, J.~C.~Helo, S.Kovalenko, and V.~E.~Lyubovitskij,
Phys. Rev. D {\bf 83}, 115015 (2011).

\bibitem{Sun}
K.~S.~Sun, T.~F.~Feng, L.~N.~Kou, F.~Sun, T.~J.~Gao, and  H.~B.~Zhang, 
Mod.~Phys.~Lett.~A {\bf 27}, 1250172 (2012).

\bibitem{Dong}
X.~X.~Dong, S.~M.~Zhao, J.~J.~Feng, G.~Z.~Ning, J.~B.~Chen, H.~B.~Zhang, and T.~F.~Feng,
Phys. Rev. D {\bf 97}, 056027 (2018).

\bibitem{BESII_emu}
J.~Z.~Bai {\it et al.} (BES Collaboration),
Phys.\ Lett.\ B {\bf 561}, 49 (2003). 

\bibitem{etau}
 M.~Ablikim {\it et al.} (BES Collaboration),
Phys. Lett. B {\bf 598}, 172 (2004).
 
\bibitem{BESIII_emu}
 M.~Ablikim {\it et al.} (BESIII Collaboration),
Phys.\ Rev.\ D {\bf 87}, 112007 (2013). 
 

\bibitem{Ablikim:2009aa}
  M.~Ablikim {\it et al.} (BESIII Collaboration),
  Nucl.\ Instrum. Methods Phys. Res., Sect. A {\bf 614}, 345 (2010).

\bibitem{Yu:IPAC2016-TUYA01}
   C.~H.~Yu {\it et al.},
  Proceedings of IPAC2016, Busan, Korea (JACoW, Geneva, Switzerland, 2016).
  
 \bibitem{Ablikim:2019hff}
M.~Ablikim \textit{et al.} (BESIII Collaboration),
Chin. Phys. C \textbf{44}, 040001 (2020).

\bibitem{etof}
 X.~Li {\it et al.}, Radiat. Detect. Technol. Methods {\bf 1}, 13 (2017);
 Y.~X.~Guo {\it et al.}, Radiat. Detect. Technol. Methods {\bf 1}, 15 (2017);
 P.~Cao {\it et al.}, Nucl.\ Instrum. Methods Phys. Res., Sect. A {\bf 953}, 163053 (2020).
 
 \bibitem{Njpsi}
M. Ablikim {\it et al.}  (BESIII Collaboration),
Chin. Phys. C {\bf 41}, 013001 (2017).
 
 \bibitem{geant4}
  S.~Agostinelli {\it et al.} (GEANT4 Collaboration),
  Nucl.\ Instrum. Methods Phys. Res., Sect. A {\bf 506}, 250 (2003).

\bibitem{ref:kkmc}
  S.~Jadach, B.~F.~L.~Ward and Z.~Was,
  Phys.\ Rev.\ D {\bf 63}, 113009 (2001);
  Comput.\ Phys.\ Commun.\  {\bf 130}, 260 (2000).  

\bibitem{ref:evtgen}
  D.~J.~Lange,
  Nucl.\ Instrum. Methods Phys. Res., Sect. A {\bf 462}, 152 (2001);
  R.~G.~Ping,
  Chin. Phys. C {\bf 32}, 599 (2008).


\bibitem{PDG2020}
P. A. Zyla {\it et al.} (Particle Data Group),
Prog. Theor. Exp. Phys. {\bf 2020}, 083C01 (2020).


\bibitem{ref:lundcharm}
  J.~C.~Chen, G.~S.~Huang, X.~R.~Qi, D.~H.~Zhang and Y.~S.~Zhu,
  Phys.\ Rev.\ D {\bf 62}, 034003 (2000);
  R.~L.~Yang, R.~G.~Ping and H.~Chen,
  Chin.\ Phys.\ Lett.\  {\bf 31}, 061301 (2014).

\bibitem{photos}
  E.~Richter-Was,
  Phys.\ Lett.\ B {\bf 303}, 163 (1993).
  



\bibitem{TAUHADNU1}
J.~H.~Kuhn and A.~Santamaria, 
Z.~Phys.~C {\bf 48}, 445 (1990).

\bibitem{TAUHADNU2}
S. Anderson {\it et al.} (CLEO Collaboration)
Phys. Rev. D \textbf{61}, 112002 (2000).


\bibitem{sys_pid_pi}
  M.~Ablikim {\it et al.} (BESIII Collaboration), 
Phys.\ Rev.\ D {\bf 83}, 112005 (2011).

\bibitem{sys_trk_pi}
  M.~Ablikim {\it et al.} (BESIII Collaboration),
  Phys.\ Rev. \ Lett.\ {\bf 107}, 092001 (2011).


\bibitem{sys_rec_pi0}
  M.~Ablikim {\it et al.} (BESIII Collaboration),
  Phys.\ Rev.\ D {\bf 81}, 052005 (2010).

\bibitem{TRolke}
W.~A.~Rolke, A.~M.~Lopez, and J.~Conrad,
Nucl. Instrum. Methods Phys. Res., Sect. A {\bf 551}, 493 (2005).

\bibitem{RooStats}
L. Moneta {\it et al.}, Proc. Sci., ACAT2010 (2010) 057.


\end{thebibliography}

\end{document}